# Arbitrary Choice of Basic Variables in Density Functional Theory. II. Illustrative Applications

Katsuhiko Higuchi

*Department of Electrical Engineering, Hiroshima University, Higashi-Hiroshima 739-8527, Japan*

Masahiko Higuchi

*Department of Physics, Shinshu University, Matsumoto 390-8621, Japan*

**Abstract**

Our recent theory (Ref. 1) enables us to choose arbitrary quantities as the basic variables of the density functional theory. In this paper we apply it to several cases. In the case where the occupation matrix of localized orbitals is chosen as a basic variable, we can obtain the single-particle equation which is equivalent to that of the LDA+U method. The theory also leads to the Hartree-Fock-Kohn-Sham equation by letting the exchange energy be a basic variable. Furthermore, if the quantity associated with the density of states near the Fermi level is chosen as a basic variable, the resulting single-particle equation includes the additional potential which could mainly modify the energy-band structures near the Fermi level.

**I. Introduction**

During past decades, the application of density functional theory (DFT)[2-7] has become the most effective method for the calculation of ground-state electronic properties of atoms, molecules and solids.[8] Exchange and correlation effects can in principle be contained in the exchange-correlation energy functional. The reproducibility of ground-state properties via Kohn-Sham (KS) orbitals and spectra is dependent on the exchange-correlation energy functional devised. We usually borrow the knowledge of the exchange-correlation energy from the homogeneous electron liquid. The knowledge is utilized not only in the local density approximation (LDA), but also in the modified schemes of the LDA such as the generalized gradient approximation (GGA),[9-11] weighted density approximation (WDA),[12-16] averaged density approximation (ADA),[12-14, 17] and so on. Of course it is one of the approved ways that the exchange-correlation energy functional $E_{xc}[\rho]$ is improved more sophisticatedly. As an instance, the use of the optimized effective potential (OEP) method has been proposed by Gross and co-workers.[18, 19]

However, there are two kinds of inconveniency in expressing the correlation effects efficiently within the framework of the conventional DFT. One is that the reproduced quantity in the reference system is the electron density alone. Another is the difficulty in devising the exchange-correlation energy functional in an appropriate form only by the use of electron density.

Let us mention the first inconvenience. The basic variable of the DFT, namely the electron density, is reproduced by means of the KS orbitals in the reference system. Every property of the ground-state is uniquely determined by the electron density through the ground-state wavefunction of the real system, but there is no insurance to reproduce the quantities other than the electron density in the reference system. Now let us illustrate the case where the spin-polarization is representative of the ground-state properties. The spin density is not necessarily reproduced in the reference system of the DFT, while both the electron density and spin density are reproduced in the reference system of the spin density functional theory (SDFT).[20-22] Of course, the spin density of the ground-state can in principle be determined in the form of $m[\rho(r)]$ via the ground-state wavefunction of the real system in the DFT. However, from the practical viewpoint the spin density should be directly calculated in the reference system. The quantities that do not correspond to basic variables, even though they are considered significant in expressing the typical properties of the ground-state, are not necessarily reproduced in the reference system.

Next we shall mention the second inconvenience. The ground-state properties can be determined by the electron density alone in the conventional DFT. The exchange-correlation energy functional is also dependent on the electron density alone, so that it is hard to devise the exchange-correlation energy functional which sufficiently reflects the correlation effects only by the use of electron density. As an example it has been shown that the local spin density approximation (LSDA) formulation is more effective in describing the ground-state properties of hydrogen atom than the LDA.[23] This is caused by the difficulty of describing the properties of the spin-polarization only through the electron density. The correlation is peculiar to the system and the appearance of the correlation varies depending on the system too. So long as we employ the conventional DFT, the highly complicated functional $E_{xc}[\rho]$ is required for the adequate description of the correlation effects.[24, 25]

For the purpose of expressing the correlation effects more directly and efficiently than the conventional DFT, we have recently proposed the extended constrained-search (ECS) formulation.[1] According to this theory, arbitrary observables can be chosen as basic variables.[1] In other words, if the ground-state properties of the system are characterized by some quantities plus the electron density, we may choose such quantities as basic variables so as to describe the correlation effects efficiently. Due to the arbitrary choice of basic variables, the ECS theory would entirely overcome the inconveniences associated with the conventional DFT, which are mentioned above. In the previous paper,[1] the SDFT and current density functional theory (CDFT)[26-33] have been revisited so as to confirm the validity of the ECS theory.

The ECS theory[1] would be promising for expressing the correlation effects of various systems efficiently, if basic variables are chosen appropriately to the systems. In this paper, we shall present several illustrative applications of the ECS theory. The various types of the correlation are described in the form of the additional potentials of single-particle equations. The organization of this paper is as follows. In section II, the single-particle equation which is equivalent to that of the LDA+U method is derived if the occupation matrix of localized orbitals is chosen as a basic variable. The ECS theory also gives the Hartree-Fock-Kohn-Sham (HF-KS) equation,[34, 35] which is shown in section III. In section IV, we discuss the cases where the quantities that are related to the density of states near the Fermi level are chosen as basic variables. The single-particle equations derived contain the additional potentials which mainly modify the energy-bands near the Fermi level. Finally in section V, we summarize and give some comments on the results of the above applications.

## II. Occupation matrix of localized orbitals as a basic variable
## - Revisiting the single-particle equation of the LDA+U method -

In the LDA+U method,[36-39] electron interactions are classified into two types on the basis of the Anderson model.[40] One is the interaction between atomic-like localized orbitals, which is supposed to be the strong on-site Coulomb interaction among them. Another is the interaction between delocalized electrons, which could be described by the ordinary LDA method. Total energy of the LDA+U method is given by the following form:[36]

$$E^{\mathrm{LSDA+U}}[\rho^\sigma, \hat{n}^\sigma] = E^{\mathrm{LSDA}}[\rho^\sigma] + E^{\mathrm{U}}[\hat{n}^\sigma] - E^{dc}[\hat{n}^\sigma], \qquad (1)$$

where $\rho^\sigma(r)$ is the electron density of spin-$\sigma$ ($\sigma = \uparrow, \downarrow$), and $E^{\mathrm{LSDA}}[\rho^\sigma]$ represents the ordinary LSDA functional. $\hat{n}^\sigma$ ($= n^\sigma_{mm'}$) is the occupation matrix of localized orbitals $\{\phi_m\}$, where $m$ is the magnetic quantum number (the other quantum numbers are abbreviated for convenience). $E^{\mathrm{U}}[\hat{n}^\sigma]$ stands for the interaction between localized electrons and is given by

$$E^{\mathrm{U}}[\hat{n}^\sigma] = \frac{1}{2} \sum_{\substack{m_1,m_2 \\ m_3,m_4}} \sum_{\sigma,\sigma'} n^\sigma_{m_1 m_2} n^{\sigma'}_{m_3 m_4} \left\{ \iint \phi^*_{m_1}(r_1) \phi^*_{m_3}(r_2) V^{ee}(r_1,r_2) \phi_{m_2}(r_1) \phi_{m_4}(r_2) \mathrm{d}r_1 \mathrm{d}r_2 \right.$$
$$\left. - \delta_{\sigma\sigma'} \iint \phi^*_{m_1}(r_1) \phi^*_{m_3}(r_2) V^{ee}(r_1,r_2) \phi_{m_4}(r_1) \phi_{m_2}(r_2) \mathrm{d}r_1 \mathrm{d}r_2 \right\}$$

(2)

where $V^{ee}$ is the effective Coulomb interaction and can be expressed in terms of the effective Slater integrals. $E^{dc}[\hat{n}^\sigma]$ in Eq. (1) is the double-counting term which approximately corresponds to the interaction between localized electrons, that is already included in the LSDA:

$$E^{dc}[\hat{n}^\sigma] = \frac{U}{2} n(n-1) - \frac{J}{2} \sum_\sigma n^\sigma(n^\sigma - 1), \tag{3}$$

where $U$ and $J$ are respectively effective Coulomb and exchange parameters, and given by using the effective Slater integrals.[36] In Eq. (3), $n = \sum_\sigma n^\sigma$ with $n^\sigma = \mathrm{Tr}(\hat{n}^\sigma)$ is the total occupation number of localized orbitals.

In the LDA+U method, the interaction energy which is not included in the LSDA scheme is

$$\Delta E[\hat{n}^\sigma] = E^{\mathrm{U}}[\hat{n}^\sigma] - E^{dc}[\hat{n}^\sigma]. \tag{4}$$

This energy can be regarded as the exchange-correlation energy which fails to be considered in the LSDA. Therefore, the exchange-correlation energy functional is reasonably given by

$$E_{xc}[\rho^\sigma, \hat{n}^\sigma] = E_{xc}^{LDA}[\rho^\sigma] + \Delta E[\hat{n}^\sigma]. \tag{5}$$

We shall apply the ECS theory to the above exchange-correlation energy functional (5). The basic variable chosen here is the occupation matrix of localized orbitals, $\hat{n}^\sigma$. Note that $\hat{n}^\sigma$ can be chosen as a basic variable since the elements of $\hat{n}^\sigma$ are calculated from the antisymmetric wavefunction. Correspondingly, the constrained-search $F[\rho^\sigma, \hat{n}^\sigma]$ is defined in accordance with the previous paper.[1] The reference system, which is described by the set of single Slater determinants, is introduced so as to reproduce basic variables. The exchange-correlation energy functional $E_{xc}[\rho^\sigma, \hat{n}^\sigma]$ is also defined as the remaining part which is given by subtracting both the kinetic energy functional of the reference system $T_s[\rho^\sigma, \hat{n}^\sigma]$ and the Hartree term from $F[\rho^\sigma, \hat{n}^\sigma]$. Equation (5) is regarded as one of the approximate forms for $E_{xc}[\rho^\sigma, \hat{n}^\sigma]$. Here we assume that the minimums exist in the definitions of $F[\rho^\sigma, \hat{n}^\sigma]$ and $T_s[\rho^\sigma, \hat{n}^\sigma]$ in the similar way to the preceding paper.[1]

In this case, basic variables other than the electron density are independent of position $r$. In order to derive the Kohn-Sham type single-particle equation, some modifications to the procedure of the previous paper are needed. The derivation is shown in the appendix. According to Eq. (A6) in the appendix, the condition under which the single-particle equation is transformed into the canonical form is

$$\sum_{m,m'} \frac{\delta E_{xc}[\rho^\sigma, \hat{n}^\sigma]}{\delta n^\sigma_{mm'}} \int \phi_k^*(r) \left( \frac{\delta n^\sigma_{mm'}}{\delta \phi_l^*(r)} \right) dr = \sum_{m,m'} \left( \frac{\delta E_{xc}[\rho^\sigma, \hat{n}^\sigma]}{\delta n^\sigma_{mm'}} \right)^* \int \left( \frac{\delta n^\sigma_{mm'}}{\delta \phi_l(r)} \right)^* \phi_k^*(r) dr, \qquad (6)$$

where $n^\sigma_{mm'}$ denote the matrix elements of $\hat{n}^\sigma$. Derivatives $\delta n^\sigma_{mm'}/\delta \phi_l^*(r)$ and $\delta n^\sigma_{mm'}/\delta \phi_l(r)$ are obtained by calculating the density matrix with the use of orbitals $\phi_l(r)$. For example, we can see the explicit expressions for the derivatives in Ref. 41, which are calculated on the basis of the LAPW method. It is easily confirmed by using such explicit expressions that Eq. (6) is satisfied. After the unitary transformation, we finally get the single-particle equations;

$$\left\{-\frac{\hbar^2 \nabla^2}{2m} + \lambda(r)\right\}\phi_k(r) + \sum_{m,m'} \frac{\delta E_{xc}[\rho^\sigma, \hat{n}^\sigma]}{\delta n_{mm'}^\sigma} \frac{\delta n_{mm'}^\sigma}{\delta \phi_k^*(r)} = \varepsilon_k \phi_k(r), \qquad (7)$$

where $\lambda(r)$ is

$$\lambda(r) = v_{ext}(r) + e^2 \int \frac{\rho_0(r')}{|r-r'|} dr' + \frac{\delta E_{xc}[\rho^\sigma, \hat{n}^\sigma]}{\delta \rho^\sigma(r)}\bigg|_{\substack{\rho^\sigma = \rho_0^\sigma \\ \hat{n}^\sigma = \hat{n}_0^\sigma}}, \qquad (8)$$

where $\rho_0(r)$ and $\hat{n}_0^\sigma$ stand for the ground-state values of basic variables, and $v_{ext}(r)$ represents the external scalar potential.

The third term on the left-hand side of Eq. (7) is the additional potential to the ordinary LSDA potential. The derivative $\delta n_{mm'}^\sigma / \delta \phi_k^*(r)$ can be obtained by using the explicit expression for $\hat{n}^\sigma$,[41] which already appeared in the evaluation of Eq. (6). Concerning the derivative $\delta E_{xc}[\rho^\sigma, \hat{n}^\sigma]/\delta n_{mm'}^\sigma$, we can rewrite it as the following expression by utilizing Eqs. (1)–(5):

$$\begin{aligned} V_{mm'}^\sigma &= \frac{\delta E_{xc}[\rho^\sigma, \hat{n}^\sigma]}{\delta n_{mm'}^\sigma} \\ &= \sum_{p,q} \sum_{\sigma'} \left\{ \iint \phi_m^*(r_1) \phi_p^*(r_2) V^{ee}(r_1, r_2) \phi_{m'}(r_1) \phi_q(r_2) dr_1 dr_2 \right. \\ &\quad \left. - \delta_{\sigma\sigma'} \iint \phi_m^*(r_1) \phi_p^*(r_2) V^{ee}(r_1, r_2) \phi_q(r_1) \phi_{m'}(r_2) dr_1 dr_2 \right\} n_{pq}^{\sigma'} \\ &\quad - \delta_{mm'} U\left(n - \frac{1}{2}\right) + \delta_{mm'} J\left(n^\sigma - \frac{1}{2}\right). \end{aligned} \qquad (9)$$

Substituting these results into Eq. (7), we get the single-particle equation which is completely equivalent to the LDA+U method.[36, 41]

## III. Exchange energy as a basic variable
## - Reconsideration of the Hartree-Fock-Kohn-Sham equation -

In this section, we shall revisit the HF-KS scheme by means of the ECS theory.[1] Let us choose the following quantity as one of basic variables:

$$E_x[\Psi] = \langle \Psi | \hat{V}_{ee} | \Psi \rangle - U_H[\rho[\Psi]] - \varepsilon_{corr.}, \tag{10}$$

where $\Psi$ is the antisymmetric wavefunction of the real system, and the first and second terms of the right-hand side are given by

$$\hat{V}_{ee} = \frac{e^2}{2} \iint dr dr' \frac{\psi^+(r)\psi^+(r')\psi(r')\psi(r)}{|r-r'|}, \tag{11}$$

$$U_H[\rho[\Psi]] = \frac{e^2}{2} \iint dr dr' \frac{\rho(r)\rho(r')}{|r-r'|}, \tag{12}$$

with

$$\rho(r) = \langle \Psi | \psi^+(r)\psi(r) | \Psi \rangle, \tag{13}$$

where $\psi(r)$ and $\psi^+(r)$ are the field operators of electrons. The last term of Eq. (10) is defined as

$$\varepsilon_{corr.} = \operatorname*{Min}_{\Psi} \langle \Psi | \hat{H} | \Psi \rangle - E_{HF}, \tag{14}$$

where $E_{HF}$ is the total energy of the system $\hat{H}$ within the Hartree-Fock approximation. Since Eq. (14) denotes the exact correlation energy, Eq. (10) corresponds to the exchange energy of the system. Note that $\varepsilon_{corr.}$ is independent of $\Psi$, i.e., it takes a constant value.

The exchange energy depends on the exchange hole which is defined as the spatial region where electrons having the same spin orientation avoid each other. Furthermore, the exchange hole can not be expressed by using the electron density alone in general cases, with the exception of the homogeneous electron liquid model. Therefore, we may assume that $E_x$ is chosen as the basic variable simultaneously with the electron density. By using the expressions of Eqs. (10) and (13), the constrained-search $F[\rho, E_x]$ can be explicitly defined under the assumption that the minimum exists, in accordance with the preceding paper.[1]

Next, we shall reproduce basic variables in the reference system which is described by the set of single Slater determinants. If we put the single Slater determinant $\Phi$ in place of $\Psi$, then Eqs. (10) and (13) are reduced to

$$E_x[\Phi] = -\frac{e^2}{2}\sum_{i,j}\iint \frac{\phi_i^*(r_1)\phi_j^*(r_2)\phi_j(r_1)\phi_i(r_2)}{|r_1 - r_2|}dr_1 dr_2 - \varepsilon_{corr.}, \tag{15}$$

$$\rho(r) = \sum_i \phi_i(r)^* \phi_i(r), \tag{16}$$

where $\phi_i(r)$ are the constituent KS orbitals of the single Slater determinant. The first term of the right-hand side in Eq. (15) has the identical form to the HF exchange energy. However, this term does not mean the exchange energy because it consists of the KS orbitals which are generally different from the HF orbitals. On the other hand, the left-hand side of Eq. (15) stands for the real exchange energy which should be reproduced.

In order to decompose the functional $F[\rho, E_x]$ into more tractable form, the kinetic energy of the reference system is introduced in the same way as the preceding paper:[1]

$$T_s[\rho, E_x] \equiv \operatorname*{Min}_{\Phi \to (\rho, E_x)} \langle \Phi | \sum_{i=1}^{N} \frac{-\hbar^2 \nabla_i^2}{2m} | \Phi \rangle. \tag{17}$$

Again, we assume that the minimum exists in Eq. (17). Using this functional, the exchange-correlation energy functional is defined as the remaining part of $F[\rho, E_x]$. Namely, we have

$$E_{xc}[\rho, E_x] \equiv F[\rho, E_x] - T_s[\rho, E_x] - U_H[\rho]. \tag{18}$$

Taking care of the fact that the basic variable (15) is independent of *r*, single-particle equations can be derived like Eqs. (A3a) and (A3b). From Eq. (A6), the condition under which the single-particle Hamiltonian is Hermitian is given by

$$\int \phi_k^*(r) \frac{\delta E_{xc}}{\delta E_x} \frac{\delta E_x}{\delta \phi_l^*(r)} dr = \int \left(\frac{\delta E_{xc}}{\delta E_x}\right)^* \left(\frac{\delta E_x}{\delta \phi_l(r)}\right)^* \phi_k^*(r) dr .$$

It is easily confirmed by using Eq. (15) that the above condition is satisfied. Therefore the single-particle equation can be transformed into the canonical form after the unitary transformation. We get

$$\left\{-\frac{\hbar^2 \nabla^2}{2m} + \lambda(r)\right\} \phi_k(r) + \frac{\delta E_{xc}[\rho, E_x]}{\delta E_x} \cdot \frac{\delta E_x[\{\phi_k\}]}{\delta \phi_k^*(r)} = \varepsilon_k \phi_k(r), \tag{19}$$

$$\lambda(r) = v_{ext}(r) + \int \frac{e^2 \rho(r')}{|r - r'|} dr' + \frac{\delta E_{xc}[\rho, E_x]}{\delta \rho(r)} . \tag{20}$$

The additional potential is the third term of Eq. (19). Substitution of Eq. (15) into the derivative $\delta E_x[\{\phi_k\}]/\delta \phi_k^*(r)$ leads to

$$\frac{\delta E_x[\{\phi_k\}]}{\delta \phi_k^*(r)} = -e^2 \sum_{i \uparrow \uparrow} \int \frac{\phi_i^*(r') \phi_k(r')}{|r - r'|} dr' \phi_i(r). \tag{21}$$

This form is identical with the exchange potential of the ordinary HF equation. The exchange-correlation energy is formally decomposed into

$$E_{xc}[\rho, E_x] = E_x + E_c[\rho]. \tag{22}$$

Here note that the functional $E_c[\rho]$ means the correlation energy which also includes the difference between the kinetic energy of the reference system and the real kinetic energy. By using Eqs. (22) and (21), the additional potential of Eq. (19) is written as

$$\frac{\delta E_{xc}[\rho, E_x]}{\delta E_x} \frac{\delta E_x[\{\phi_k\}]}{\delta \phi_k^*(r)} = -e^2 \sum_{i\uparrow\uparrow} \int \frac{\phi_i^*(r')\phi_k(r')}{|r - r'|} dr' \phi_i(r). \tag{23}$$

On the other hand, the potential $\lambda(r)$ contains the correlation part $\delta E_{xc}[\rho, E_x]/\delta\rho(r) = \delta E_c[\rho]/\delta\rho(r)$. Therefore, the resulting single-particle equation has the Hartree-Fock type exchange potential plus the correlation potential of the ordinary DFT. It is sometimes called Hartree-Fock-Kohn-Sham equation.[34, 35] Of course, if the correlation energy is neglected in Eq. (22), then we get the single-particle equation which is equal to the Hartree-Fock equation.

Göring, Levy and co-workers have derived the HF-KS equation in the manner different from the present scheme.[42-44] They first introduced the functional $F^S[\rho]$ which is defined as the expectation value of kinetic and electron-electron interaction operators with respect to the minimizing single Slater determinant yielding the ground-state value of the electron density. By employing the arbitrariness concerning the decomposition of the functional $F[\rho]$ in the reference system, $F[\rho]$ is decomposed into two parts, i.e., $F^S[\rho]$ and the remaining part. If the remaining part is supposed to be the correlation energy functional, then the HF-KS equation can be derived.[42-44]

In deriving the HF-KS equation, Göring *et al.* utilize the arbitrariness on the decomposition of functional $F[\rho]$, while we utilize the arbitrariness on the choice of basic variables. The merit of our theory is to get the ground-state value of $E_x$ or $E_x + \varepsilon_{corr.}$ through Eq. (15), together with the electron density.

### IV. Basic variables associated with the density of states near the Fermi level

It is hard to ignore the correlation effects in systems where the density of states near the Fermi level shows the high value. This can be reasonably understood if one recalls the correlation energy $\varepsilon_{corr.}$ of the configuration interaction (CI) method.[45-48] The doubly excited CI (DCI) is sufficient for the estimation of $\varepsilon_{corr.}$, because the contributions of single excitations to $\varepsilon_{corr.}$ are the fourth and higher-order perturbation terms.[46, 47] The approximate form of $\varepsilon_{corr.}$ is given by[46]

$$\varepsilon_{corr.} \cong -\sum_{i,j}^{occ.} \sum_{a,b}^{unocc.} \frac{\langle \Phi_{HF} | \hat{H} | \Phi_{ij}^{ab} \rangle \langle \Phi_{ij}^{ab} | \hat{H} | \Phi_{HF} \rangle}{\langle \Phi_{ij}^{ab} | \hat{H} - E_{HF} | \Phi_{ij}^{ab} \rangle} \tag{24}$$

where $\Phi_{HF}$ denotes the Hartree-Fock determinant which is formed from $N$ lowest energy orbitals. $E_{HF}$ stands for the ground-state energy of the Hartree-Fock approximation, which is given by $E_{HF} = \langle \Phi_{HF} | \hat{H} | \Phi_{HF} \rangle$. $\Phi_{ij}^{ab}$ is the doubly excited determinant which differs from $\Phi_{HF}$ in replacing occupied orbitals $\chi_i$ and $\chi_j$ by unoccupied orbitals $\chi_a$ and $\chi_b$, respectively.[46]

One of the important features of this correlation energy is that it prompts the orbital mixing between occupied and unoccupied states near the Fermi level. The gain of the correlation energy increases with the density of states near the Fermi level. In other words, the correlation energy increases as occupied and unoccupied states come close to each other. Thus, the correlation energy is dependent on the density of states in the vicinity of the Fermi level. In this section we shall consider two cases where the quantities which are related to the density of states near the Fermi level are chosen as basic variables.   At the first half of this section, we deal with the case in which the sum of the electron density within the restricted energy region is chosen as a basic variable.   Later this section, the local density of states at the Fermi level is chosen as a basic variable.

**IV-a.   Sum of the electron density within the restricted energy region**

In order to reflect the above features of the correlation energy explicitly, we shall choose as the basic variable the sum of the local density of states over the energy region centered at the Fermi level $\varepsilon_F$.   Suppose that the energy region is

$$\varepsilon_F - \Delta < \varepsilon < \varepsilon_F + \Delta, \qquad (25)$$

where $\Delta$ is the appropriate energy parameter.   The local density of states at some energy $\varepsilon$ is defined as the imaginary part of the retarded Green's function, i.e.,

$$d(\boldsymbol{r},\omega) = -\frac{1}{\pi} \operatorname{Im} G^R(\boldsymbol{r},\boldsymbol{r};\omega), \qquad (26)$$

where $\omega = \varepsilon/\hbar$.   Note that the local density of states $d(\boldsymbol{r},\omega)$ results in the density of states at $\varepsilon$ by integrating it with respect to $\boldsymbol{r}$.   Combining (25) and (26), the basic variable chosen here is

$$\begin{aligned} D(\boldsymbol{r},\omega_F) &= \int_{\varepsilon_F - \Delta}^{\varepsilon_F + \Delta} d(\boldsymbol{r},\omega) d\varepsilon \\ &= -\frac{1}{\pi} \int_{\varepsilon_F - \Delta}^{\varepsilon_F + \Delta} \operatorname{Im} G^R(\boldsymbol{r},\boldsymbol{r};\omega) d\varepsilon, \end{aligned} \qquad (27)$$

where $\omega_F = \varepsilon_F/\hbar$.   According to the preceding paper,[1] the $F$-functional, i.e., $F[\rho, D]$ can be defined under the assumption that the set of antisymmetric wavefunctions yielding $\rho(\boldsymbol{r})$ and $D(\boldsymbol{r},\omega_F)$ is weak closed and not empty.   The extended Hohenberg-Kohn theorem[1] holds for basic variables $\rho(r)$ and $D(r,\omega_F)$.

In order to reproduce basic variables by using single-particle wavefunctions and spectra, we shall adopt as the reference system the set of wavefunctions which are in the form of simple products of $N$ single-particle wavefunctions. Namely, the element of the set is written by

$$\Xi[\{\varphi_i\}] = \varphi_{\lambda_1}(r_1)\varphi_{\lambda_{2_1}}(r_2)\varphi_{\lambda_3}(r_3)\cdots\varphi_{\lambda_N}(r_N). \tag{28}$$

For the purpose of decomposing $F[\rho,D]$, the following functional is newly introduced:

$$\begin{aligned}\widetilde{T}_s[\rho,D] &= \underset{\Xi[\{\varphi_i\}]\mapsto(\rho,D)}{\text{Min}} \langle \Xi[\{\varphi_i\}] | \sum_{j=1}^{N} -\frac{\hbar^2 \nabla_j^2}{2m} | \Xi[\{\varphi_i\}] \rangle \\ &= \underset{\Xi[\{\varphi_i\}]\mapsto(\rho,D)}{\text{Min}} \sum_{i=\lambda_1}^{\lambda_N} \langle \varphi_i | -\frac{\hbar^2 \nabla^2}{2m} | \varphi_i \rangle,\end{aligned} \tag{29}$$

where $\Xi[\{\varphi_i\}] \to (\rho,D)$ indicates that the search is constrained among all $\Xi[\{\varphi_i\}]$ which yield the prescribed $\rho(r)$ and $D(r,\omega_F)$. Here we assume again that the set of (28) yielding $\rho(r)$ and $D(r,\omega_F)$ is weak closed and not empty. The functional $\widetilde{T}_s[\rho,D]$ corresponds to the kinetic energy of the peculiar reference system which is mentioned above. Note that the orthogonality of $N$ orbitals $\{\varphi_i\}$ is not necessarily demanded.

The basic variables should be reproduced by taking the expectation values of their operators with respect to $\Xi[\{\varphi_i\}]$. The electron density is given by

$$\rho(r) = \sum_i |\varphi_i(r)|^2. \tag{30}$$

On the other hand, we shall assume the reproduced form of $D(r,\omega_F)$. The retarded Green's function is generally written by using quasiparticle wavefunctions and spectra.[49] If quasiparticle wavefunctions and spectra are respectively replaced with the KS orbitals and spectra, and if the imaginary parts of quasiparticle spectra are supposed to be constant $a$, then the local density of states is written as

$$d(\mathbf{r},\omega) = \sum_j \varphi_j(\mathbf{r})^* \varphi_j(\mathbf{r}) \frac{1}{a\sqrt{\pi}} e^{-\frac{(\omega-\varepsilon_j/\hbar)^2}{a^2}} . \tag{31}$$

Utilizing Eq. (31), we suppose that the reproduced expression for $D(\mathbf{r},\omega_F)$ is given by

$$D(\mathbf{r},\omega_F) = \sum_j \int_{\varepsilon_F-\Delta}^{\varepsilon_F+\Delta} d\varepsilon \varphi_j(\mathbf{r})^* \varphi_j(\mathbf{r}) \frac{1}{a\sqrt{\pi}} e^{-\frac{(\omega-\varepsilon_j/\hbar)^2}{a^2}} . \tag{32}$$

The derivation procedure of the single-particle equation is a little different from the previous one (Eqs. (4-9), (4-5) and (4-6) in Ref. 1), since the orthogonality is not imposed on $N$ orbitals $\{\varphi_i\}$ in Eq. (29). Correspondingly we do not need the unitary transformation of the single-particle equation. Taking care of the above points, we obtain the self-consistent single-particle equations:

$$\left\{ -\frac{\hbar^2 \nabla^2}{2m} + \lambda(\mathbf{r}) \right\} \varphi_k(\mathbf{r}) + \int \mu(\mathbf{r}') \frac{\delta D(\mathbf{r}',\omega_F)}{\delta \varphi_k^*(\mathbf{r})} d\mathbf{r}' = \varepsilon_k \varphi_k(\mathbf{r}), \tag{33}$$

where $\lambda(\mathbf{r})$ and $\mu(\mathbf{r})$ are respectively given by

$$\lambda(\mathbf{r}) = v_{ext}(\mathbf{r}) + e^2 \int \frac{\rho_0(\mathbf{r}')}{|\mathbf{r}-\mathbf{r}'|} d\mathbf{r}' + \frac{\delta E_{xc}[\rho,D]}{\delta \rho(\mathbf{r})} \bigg|_{\substack{\rho=\rho_0 \\ D=D_0}}, \tag{34}$$

and

$$\mu(\mathbf{r}) = \frac{\delta E_{xc}[\rho,D]}{\delta D} \bigg|_{\substack{\rho=\rho_0 \\ D=D_0}} . \tag{35}$$

In Eqs. (34) and (35), $\rho_0(r)$ and $D_0(r,\omega_F)$ stand for the ground-state values of basic variables.

In order to investigate the features of the single-particle equation, let us consider the derivative $\delta D(r',\omega_F)/\delta\varphi_k^*(r)$ which appears in Eq. (33). Using Eq. (32), the derivative is written as

$$\frac{\delta D(r',\omega_F)}{\delta\varphi_k^*(r)} = \delta(r-r')\varphi_k(r')\frac{1}{a\sqrt{\pi}}\int_{\varepsilon_F-\Delta}^{\varepsilon_F+\Delta} e^{-\frac{(\varepsilon_k-\varepsilon)^2}{a^2}}d\varepsilon \\ + \sum_i \frac{1}{a\sqrt{\pi}}\varphi_i^*(r')\varphi_i(r')\frac{\delta}{\delta\varphi_k^*(r)}\int_{\varepsilon_F-\Delta}^{\varepsilon_F+\Delta} e^{-\frac{(\varepsilon_i-\varepsilon)^2}{a^2}}d\varepsilon. \quad (36)$$

By means of the chain rule for the functional derivatives, the second term in Eq. (36) is

$$\frac{\delta}{\delta\varphi_k^*(r)}\int_{\varepsilon_F-\Delta}^{\varepsilon_F+\Delta} e^{-\frac{(\varepsilon_i-\varepsilon)^2}{a^2}}d\varepsilon = \sum_j \frac{\delta\varepsilon_j}{\delta\varphi_k^*(r)}\frac{\partial}{\partial\varepsilon_j}\int_{\varepsilon_F-\Delta}^{\varepsilon_F+\Delta} e^{-\frac{(\varepsilon_i-\varepsilon)^2}{a^2}}d\varepsilon. \quad (37)$$

Here the functional derivative with respect to $\varepsilon_j$ is replaced by the ordinary derivative in Eq. (37), because the integral of Eq. (37) is the ordinary function of $\varepsilon_j$. We take the ordinary derivative in Eq. (37), giving the attention to the fact that $\varepsilon_F$ is a constant value which is identical with the highest occupied KS spectrum. Then, Eq. (37) is

$$\frac{\delta}{\delta\varphi_k^*(r)}\int_{\varepsilon_F-\Delta}^{\varepsilon_F+\Delta} e^{-\frac{(\varepsilon_i-\varepsilon)^2}{a^2}}d\varepsilon = -\frac{\delta\varepsilon_i}{\delta\varphi_k^*(r)}\left\{e^{-\frac{(\varepsilon_i-\varepsilon_F-\Delta)^2}{a^2}} - e^{-\frac{(\varepsilon_i-\varepsilon_F+\Delta)^2}{a^2}}\right\}. \quad (38)$$

The derivative $\delta\varepsilon_i/\delta\varphi_k^*(r)$ can be approximated in a more simplified form. The third term of the left-hand side of Eq. (33) explicitly includes the state $\varphi_k(r)$ which is the solution of the equation.

Correspondingly the single-particle Hamiltonian of Eq. (33) directly depends on the solution $\varphi_k(\mathbf{r})$, so that the predominant effect of $\delta\varepsilon_i / \delta\varphi_k^*(\mathbf{r})$ would be caused from the case of $i = k$. Under this approximation, Eq. (38) is written as

$$\frac{\delta}{\delta\varphi_k^*(\mathbf{r})} \int_{\varepsilon_F-\Delta}^{\varepsilon_F+\Delta} e^{-\frac{(\varepsilon_i-\varepsilon)^2}{a^2}} d\varepsilon \approx -\delta_{ki} \frac{\delta\varepsilon_i}{\delta\varphi_k^*(\mathbf{r})} \left\{ e^{-\frac{(\varepsilon_i-\varepsilon_F-\Delta)^2}{a^2}} - e^{-\frac{(\varepsilon_i-\varepsilon_F+\Delta)^2}{a^2}} \right\}. \tag{39}$$

Substituting Eq. (39) into Eq. (36), the single-particle equation (33) is simplified into

$$\left\{ -\frac{\hbar^2 \nabla^2}{2m} + \lambda(\mathbf{r}) \right\} \varphi_k(\mathbf{r}) + \mu(\mathbf{r}) \frac{1}{a\sqrt{\pi}} \int_{\varepsilon_F-\Delta}^{\varepsilon_F+\Delta} e^{-\frac{(\varepsilon_k-\varepsilon)^2}{a^2}} d\varepsilon\, \varphi_k(\mathbf{r})$$
$$+ \int \mu(\mathbf{r}')\varphi_k^*(\mathbf{r}')\varphi_k(\mathbf{r}')d\mathbf{r}' \frac{1}{a\sqrt{\pi}} \frac{\delta\varepsilon_k}{\delta\varphi_k^*(\mathbf{r})} \left\{ e^{-\frac{(\varepsilon_k-\varepsilon_F-\Delta)^2}{a^2}} - e^{-\frac{(\varepsilon_k-\varepsilon_F+\Delta)^2}{a^2}} \right\} = \varepsilon_k \varphi_k(\mathbf{r}).$$

(40)

Let us consider the features of the above single-particle equation. Compared to the KS equation of the conventional DFT, additional potentials are the third and fourth terms of the left-hand side. The integrand of the third term is the delta-type function which is centered at $\varepsilon_k$. The third term is proportional to the area under this delta-type function between $\varepsilon_F - \Delta$ and $\varepsilon_F + \Delta$. As the state $\varepsilon_k$ comes close to the Fermi level $\varepsilon_F$, the area increases. Therefore, the third term chiefly modifies the states which are located in the vicinity of the Fermi level. Also, the fourth term has an effect on the states only in the vicinity of the Fermi level. The energy term in the parenthesis of the fourth term has the positive or negative value depending on the position of $\varepsilon_k$. Namely, if $\varepsilon_k$ is located below the Fermi level it takes the positive value, and conversely if $\varepsilon_k$ is located above the Fermi level it takes the negative value. The fourth term would enlarge or reduce the width of energy-band structures near the Fermi level. The potential which sign depends on whether the state $\varepsilon_k$ is occupied or unoccupied can also be seen in the LDA+U method and GW approximation.[50, 51] Of course, according to the derivation processes of the above effective potential and the single-particle equation,[1] Eq. (40) is valid only for the occupied states in the same way as the KS equation of the conventional DFT.   However, it is interesting that the fourth term has the possibility of modifying the width of the energy bands near the Fermi level, like the LDA+U and GW methods.

**IV-b.   Local density of states at the Fermi level**

As already mentioned, the correlation energy (24) increases with the density of states near the Fermi level.   On the basis of this expression, the approximate model for the correlation energy functional can be discussed.   As an example, we shall consider the model where occupied and unoccupied energies in the denominator are approximated into the representative energies that are the centers of gravity for occupied and unoccupied bands respectively.[52]   This model maintains the above feature of the correlation energy.   In such a model, the representative energies for occupied and unoccupied bands become closer to the Fermi level as the density of states at the Fermi level increases.   Namely, the correlation energy increases with the density of states at the Fermi level. Therefore the density of states at the Fermi level seems to be one of the significant candidates of basic variables.   In this subsection, we shall choose the local density of states at the Fermi level as one of basic variables and derive the single-particle equation with the aid of the results of the preceding paper.[1]

We suppose that the reproduced form of the local density of states at the Fermi level is given by Eq. (31);

$$d(\bm{r},\omega_F) = \sum_j \varphi_j(\bm{r})^* \varphi_j(\bm{r}) \frac{1}{a\sqrt{\pi}} e^{-\frac{(\omega_F - \varepsilon_j/\hbar)^2}{a^2}} . \tag{41}$$

Under the choice of $d(\bm{r},\omega_F)$ and $\rho(\bm{r})$ as basic variables, self-consistent single-particle equations can be derived similarly to the previous subsection.   The additional potential which does not exist in the KS equation of the conventional DFT is given by

$$\int \mu(\mathbf{r}') \frac{\delta d(\mathbf{r}', \omega_F)}{\delta \varphi_k^*(\mathbf{r})} d\mathbf{r}' \tag{42}$$

with

$$\mu(\mathbf{r}) = \left. \frac{\delta E_{xc}[\rho, d]}{\delta d} \right|_{\substack{\rho=\rho_0 \\ d=d_0}}, \tag{43}$$

where $\rho_0(\mathbf{r})$ and $d_0(\mathbf{r}, \omega_F)$ represent the ground-state values of basic variables. Let us consider the derivative $\delta d(\mathbf{r}', \omega_F) / \delta \varphi_k^*(\mathbf{r})$ which appears in Eq. (42). Applying the same approximation which was employed in the derivation of Eq. (39), $\delta d(\mathbf{r}', \omega_F) / \delta \varphi_k^*(\mathbf{r})$ is written as

$$\frac{\delta d(\mathbf{r}', \omega_F)}{\delta \varphi_k^*(\mathbf{r})} = \delta(\mathbf{r} - \mathbf{r}') \varphi_k(\mathbf{r}') \frac{1}{a\sqrt{\pi}} e^{-\frac{(\varepsilon_k - \varepsilon_F)^2}{a^2}} \\ + \varphi_k^*(\mathbf{r}') \varphi_k(\mathbf{r}') \frac{2(\varepsilon_F - \varepsilon_k)}{a^3 \sqrt{\pi}} e^{-\frac{(\varepsilon_k - \varepsilon_F)^2}{a^2}} \frac{\delta \varepsilon_k}{\delta \varphi_k^*(\mathbf{r})}. \tag{44}$$

The resultant potential has the similar properties to those of the additional potentials of Eq. (40). The first term of Eq. (44) affects only energy-bands in the vicinity of the Fermi level, which corresponds to the third term of Eq. (40). The second term also influences the modification of states near the Fermi level. Furthermore the sign of the second term varies depending on the position of $\varepsilon_k$ in the similar way to the fourth term of Eq. (40).

From the viewpoint of practical calculations, it is difficult to deal with the derivative $\delta \varepsilon_k / \delta \varphi_k^*(\mathbf{r})$ which appears in the second term of Eq. (44). However, if neglecting the orbital dependency of the effective potential of the single-particle equation, we can utilize the approximate expression which is based on the first-order perturbation theory, i.e.,

$$\frac{\delta \varepsilon_k}{\delta \varphi_k^*(r)} = \varepsilon_k \varphi_k(r) + \int \left\{ \sum_{j \neq k} \frac{\varphi_j(r')\varphi_k^*(r')\varphi_j^*(r)}{\varepsilon_k - \varepsilon_j} \right\}^{-1} \varphi_k(r')\varphi_k(r')dr'. \qquad (45)$$

## V. Summary and Discussions

In this paper we give several types of single-particle equations as the applications of the ECS theory. Validity of the theory is confirmed by revisiting the single-particle equation of the LDA+U method and Hartree-Fock-Kohn-Sham equation in sections II and III, respectively, just as the SDFT and CDFT formulations have been derived on the basis of the ECS theory in the preceding paper.[1]

The ECS theory seems to be promising for the electronic structures of f-electron materials. The f-electron materials which belong to the heavy Fermion system show peculiar electronic characteristics.[53, 54] For instance, they exhibit the extremely large electronic specific heat coefficient, the magnitude of which is two to three orders larger than that of usual metals. This means that the effective mass becomes far heavier than the free-electron mass. Such peculiar features are sure to be caused by the electron correlations in which f-electrons play a vital part.

The density of states at the Fermi level is comparatively high for the f-electron materials which belong to the valence fluctuation regime.[55] It is due to the concentration of f-bands in the vicinity of the Fermi level. The valence fluctuation regime is considered to be smoothly connected to the itinerant f-band model.[55-57] The f-bands near the Fermi level would determine the peculiar behaviors of the valence fluctuation regime. In section IV, single-particle equations are derived by choosing the quantities related to the density of states near the Fermi level as basic variables. The resulting single-particle equations include the additional potentials which mainly modify the energy-bands near the Fermi level. If such additional potentials narrow the width of f-bands near the Fermi level, our scheme shown in section IV has the possibility of describing the electronic structures of the valence fluctuation regime more appropriately than the conventional LDA.

On the other hand, in the f-electron materials of the Kondo regime no f-bands are observed at the Fermi level and the Fermi surface is essentially the same as that of the corresponding reference material, for example, La compounds for the Ce-based heavy Fermion compound.[55-57] In these materials, f-electrons have been successfully treated with the models which presuppose the strong on-site Coulomb interaction among them.[53, 55] In section II, the single-particle equation of the LDA+U method can be revisited by the ECS theory. This fact not only implies the validity of the ECS theory, but also leads to the following significant applications. If electron correlations of the system may be recognized well on the basis of some model Hamiltonian, we may derive the single-particle equation by taking as one of basic variables the quantity which is included in such a model. Namely, the ECS theory allows us to incorporate the properties of the model Hamiltonian into the single-particle problem. In Sec. II, we illustrate the LDA+U scheme as one of the cases offering the above possibility.

**Acknowledgements**

The authors are grateful to Akira Hasegawa for continual discussions on energy-band theory. One of the authors (M. H.) would like to express his thanks to the Alexander von Humboldt Foundation for facilitating both the stay and the research in Dresden.

**Appendix: Self-consistent single-particle equations with constant basic variables**

In this appendix, self-consistent single-particle equations are derived in the case where basic variables chosen are independent of *r*. This is the simple extension of Ref. 1, in which self-consistent single-particle equations are derived for *r*-dependent basic variables.

Suppose that quantities $C_1$, $C_2$, …, $C_M$ are chosen as basic variables in addition to the electron density $\rho(r)$. In Ref. 1, basic variables are assumed to be chosen so that the minimums exist in Eqs. (2-3) and (3-3) of Ref. 1. In the following we also assume that the set of $\rho(r)$, $C_1$, …, and $C_M$ is of that type. The extended Hohenberg-Kohn theorem can be proven in the same way as the proof in Ref. 1. The kinetic energy functional of the reference system is also introduced so as to obtain the practical scheme for calculating ground-state properties. The kinetic energy functional is defined by

$$T_s[\rho, C_1, \cdots, C_M] \equiv \underset{\{\phi_i\} \to (\rho, C_1, \cdots, C_M)}{\mathrm{Min}} \sum_{i=1}^{N} \langle \phi_i | -\frac{\hbar^2 \nabla^2}{2m} | \phi_i \rangle, \quad \int \rho \, dr = N, \quad \int \left| \nabla \rho^{1/2} \right|^2 dr < \infty. \quad (A1)$$

The notation $\{\phi_i\}$ stands for a set of *N* orbitals, and $\{\phi_i\} \to (\rho, C_1, \cdots, C_M)$ means that the minimization is performed among the sets of *N* orbitals which are orthonormal and yield a given set of $\rho(r)$, $C_1$, …, and $C_M$. Thus, one obtains the minimizing *N* orbitals by searching the minimum value of $\sum_{i=1}^{N} \langle \phi_i | -\frac{\hbar^2 \nabla^2}{2m} | \phi_i \rangle$ under the conditions that the orbitals are orthonormal and yield the given set $(\rho, C_1, \cdots, C_M)$. This problem is translated into the minimization of the following functional:

$$\Omega[\{\phi_i\}] \equiv \sum_{i=1}^{N} \int \phi_i^+(r) \left( -\frac{\hbar^2 \nabla^2}{2m} \right) \phi_i(r) \, dr + \int \lambda(r) \left\{ \sum_{i=1}^{N} \phi_i^+(r) \phi_i(r) - \rho(r) \right\} dr$$
$$+ \sum_{m=1}^{M} \mu_m \{C_m[\{\phi_i(r)\}] - C_m\} - \sum_{i,j=1}^{N} \varepsilon_{ij} \left\{ \int \phi_i^+(r) \phi_j(r) \, dr - \delta_{i,j} \right\} \quad (A2)$$

Here $\lambda(r)$, $\mu_m$ and $\varepsilon_{ij}$ are Lagrange multiplier function and multipliers, respectively. $C_m[\{\phi_i(r)\}]$ in Eq. (A2) indicates that $C_m$ is given in terms of $N$ orbitals. The minimizing condition of $\Omega[\{\phi_i\}]$ leads to the necessary conditions on the minimizing $N$ orbitals. We get

$$-\frac{\hbar^2 \nabla^2}{2m}\phi_k(r) + \lambda(r)\phi_k(r) + \sum_{m=1}^{M}\mu_m\left(\frac{\delta C_m[\{\phi_i(r)\}]}{\delta \phi_k^+(r)}\right) = \sum_{j=1}^{N}\varepsilon_{kj}\phi_j(r), \qquad (A3a)$$

$$-\frac{\hbar^2 \nabla^2}{2m}\phi_k^+(r) + \lambda(r)\phi_k^+(r) + \sum_{m=1}^{M}\mu_m\left(\frac{\delta C_m[\{\phi_i(r)\}]}{\delta \phi_k(r)}\right) = \sum_{i=1}^{N}\varepsilon_{ik}\phi_i^+(r). \qquad (A3b)$$

$\lambda(r)$ and $\mu_m$ should be determined by using the extended Hohenberg-Kohn theorem. If the ground-state values of $\rho(r)$, $C_1$, …, and $C_M$ are respectively denoted by $\rho_0(r)$, $C_1^0$, …, and $C_M^0$, then $\lambda(r)$ and $\mu_m$ for the ground-state values are given by

$$\lambda[\rho_0, C_1^0, \cdots, C_M^0] = v_{ext}(r) + \int \frac{e^2 \rho_0(r')}{|r-r'|}dr' + \left.\frac{\delta E_{xc}[\rho, C_1, \cdots, C_M]}{\delta \rho(r)}\right|_{\substack{\rho=\rho_0 \\ C_m=C_m^0}}, \qquad (A4)$$

$$\mu_m[\rho_0, C_1^0, \cdots, C_M^0] = \left.\frac{\delta E_{xc}[\rho, C_1, \cdots, C_M]}{\delta X(r)}\right|_{\substack{\rho=\rho_0 \\ C_m=C_m^0}}, \qquad (A5)$$

where $E_{xc}[\rho, C_1, \cdots, C_M]$ is the exchange-correlation energy functional. Furthermore, Eqs. (A3a) and (A3b) can be changed by a unitary transformation to the canonical forms if quantities $C_1$, $C_2$, …, $C_M$ satisfy the following relation:

$$\sum_{m=1}^{M}\mu_m\int \phi_k^+(r)\left(\frac{\delta C_m[\{\phi_i(r)\}]}{\delta \phi_l^+(r)}\right)dr = \sum_{m=1}^{M}\mu_m^*\int \left(\frac{\delta C_m[\{\phi_i(r)\}]}{\delta \phi_l(r)}\right)^* \phi_k^*(r)dr. \qquad (A6)$$

This relation is regarded as the condition under which the single-particle Hamiltonian of the left-hand side of Eq. (A3a) is Hermitian. The canonical form of the single-particle equation is written as

$$-\frac{\hbar^2 \nabla^2}{2m}\phi_k(\boldsymbol{r}) + \lambda(\boldsymbol{r})\phi_k(\boldsymbol{r}) + \sum_{m=1}^{M}\mu_m\left(\frac{\delta C_m[\{\phi_i(\boldsymbol{r})\}]}{\delta \phi_k^+(\boldsymbol{r})}\right) = \varepsilon_k \phi_k(\boldsymbol{r}). \qquad (A7)$$

Eqs. (A7), (A4) and (A5) are self-consistent single-particle equations for basic variables $\rho(\boldsymbol{r})$, $C_1$, …, and $C_M$. These equations correspond to Eqs. (4-9), (4-5) and (4-6) in Ref. 1.